\documentstyle[twoside,fleqn,espcrc2]{article}
\begin{document}
\pagestyle{myheadings}
\markright{UPR--714--T}
\date{January 1997}
\title{Properties of Black Holes in Toroidally Compactified String Theory 
}
\author{Mirjam Cveti\v c
\address{University of Pennsylvania,\\
Department of Physics and Astronomy,\\
Philadelphia, PA 19104, U.S.A.}
\thanks{Compilation of talks given at SUSY'96,
CERN Workshop on Duality in String Theory II,  Strings'96 and Buckow'96.}%
}
\begin{abstract}
{We review the  macroscopic and microscopic properties of black holes 
of toroidally compactified  heterotic and Type II string theory in
dimensions $4\le D\le 9$. 
General charged rotating black hole solutions are obtained by acting on
 a generating solution with classical duality
symmetries. In $D=4$, $D=5$ and $6\le D \le 9$, the generating solution
for {\it both}  toroidally compactified Type II and heterotic strings is
 specified by the ADM mass,  $[{{D-1}\over 2}]$-angular momentum
components
and  five, three  and two charges, respectively. We give 
  the Bekenstein-Hawking entropy for  these solutions, 
  address the  BPS-saturated limit and compare the results to calculations of 
   the microscopic entropy both in the NS-NS sector and the R-R sector of
    the theory.
We also interpret such black hole  solutions as dimensionally reduced
 intersecting p-branes of  M-theory.}
\end{abstract}
\maketitle

\section{Introduction}

\setcounter{footnote}{0}
Black holes play an important role in string theory and
recent developments  (for a   review, see {\it e.g}, 
 \cite{Horowitz,Maldacena}) have shown that string theory makes
  it possible  to address    their microscopic  properties,
in particular the statistical origin of  black hole
 entropy, the radiation rates and possibly  issues of information loss for  black
hole
 physics.

The starting point in such investigations is the classical black hole solution
and one aim of this  contribution is to review  briefly the properties of a
 {\it general} class of these  solutions in toroidally compactified
heterotic and Type II string theories, {\it i.e.} for string vacua
 with enough (super)symmetries and thus well understood moduli spaces
 (This excludes notable  examples of  black hole solutions 
for $N=2$ string vacua; see {\it e.g.}, Ref.  \cite{Dieter} and 
references therein.).

In Section 2 we  review the procedure to obtain the explicit form for
  general rotating black holes in dimensions $4\le D\le 9$ 
  and  in Section 3  we  concentrate  on the structure of their Bekenstein-Hawking  (BH) entropy.
 In Section 4 we review the progress made in identifying microscopic 
degrees of freedom of such black holes in order to  obtain  a microscopic
 (statistical) interpretation of  the  BH entropy both  for a class of
  black holes with charges arising in the  Neveu-Schwarz- Neveu-Schwarz (NS-NS)
   as well as  dramatic progress made in  the  Ramond-Ramond (R-R) sector of string
theory.
In Section 5 we briefly discuss the  higher-dimensional interpretation of these 
black hole solutions as intersecting p-branes of the M-theory.

\section{Classical Black Hole Solutions}

The aim is to find  general axi-symmetric solutions of  the bosonic 
sector of the effective Lagrangian of toroidally compactified string theory.
Since the bosonic sector includes the graviton, $U(1)$ gauge fields, 
 as well as 
massless scalar fields,  the axi-symmetric solutions correspond to the 
{\it dilatonic charged rotating black hole solutions}. 
Since for such configurations  the scalar fields vary  with the spatial 
direction,  they affect their  space-time structure and thus
such solutions need not correspond to the  solutions with regular horizons. 
It turns out, however, that when  enough charges are turned on the  scalar
fields ``slow-down'' enough that the solutions   effectively correspond 
to those  with the global space-time of  the Kerr-Newmann  [Reissner-Nordstr\" om]
black holes, {\it i.e.}   global space-time is that of charged rotating [static]
black holes of ordinary Maxwell-Einstein gravity. 

According to the ``no-hair theorem'' such  black holes are specified by 
 the ADM  mass,   $[{{D-1}\over
2}]$-components of angular
momentum
and  the  number of  allowed charge parameters associated with $U(1)$ 
gauge symmetry. {\it E.g.}, for toroidally compactified heterotic strings
in $D=4$, the general black hole solution is specified by the ADM mass,
one component of the angular momentum, and  28 electric and 28 magnetic
charges associated with 28 $U(1)$ gauge-fields.

Among such black hole solutions those which 
saturate the Bogomol'nyi-Gibbons-Hull bound \cite{GH},  
are special. They correspond to configurations
with a number of  preserved supersymmetries and for enough ($N\ge 4$ in $D=4$)
supersymmetries their classical properties are protected from quantum
corrections. Thus we  can trust their  classical properties in 
the strong coupling regime as well. Such solutions play a crucial
role \cite{HT} in  establishing duality symmetries between different string
vacua. At special points of moduli space, where they can become massless, 
they  may enhance gauge symmetry \cite{Strominger,HTII} as well as
supersymmetry \cite{CYplb,Vafa}. When their space-time is that of 
black holes with regular horizons they  turn out to play a crucial 
role as solutions whose microscopic entropy 
can be identified with the BH entropy.

\subsection{General Solution from  Generating Solution }

The most general black hole,
 compatible with the ``no-hair theorem'',
 is  obtained by acting on   the generating solution  with 
 classical $U$- ($S$-, \hbox{$T$-)} duality  transformations (See Tables 1 and 2) .
These are symmetries of the supergravity equations of motion, and so generate
new solutions from old.
They  do
not change
the $D$-dimensional Einstein-frame metric but do change the charges and scalar
fields. One  first  considers transformations,  belonging
 to the maximal compact subgroup $C_U$ of duality 
 transformations (See Tables 1 and 2),  which  preserve the canonical
asymptotic values of the scalar fields and show that all charges are generated
in this way.
Another  duality transformation can be used to change the
asymptotic values of the scalar fields. Ultimately, 
the solution could be cast in the manifestly duality invariant form.

\subsubsection{Duality Symmetries}

The low-energy effective action for
the Type II string  (or M-theory), toroidally compactified
 to $D$-dimensions,  is
the maximal supergravity theory, which has a continuous duality symmetry $U$ of
its equations of motion \cite{CJ}  (see Table 1, first column). $U$-duality   has a
maximal compact subgroup $C_U$  (second  column in  Table 1).  In the  quantum
theory
the continuous classical symmetry
is broken to  the discrete  subgroup ${ Q}_{ U}$  \cite{HT} 
(third  column in Table 1) of  $U$-duality symmetry.

Toroidally compactified heterotic strings in $D$-dimensions 
 inherit the classical symmetries of the
even self-dual lattice of the (bosonic) gauge sector of the heterotic
string and that of $T^{10-D}$ torus (the first
column in Table 2),  referred to as  continuous $T$- and $S$-duality
symmetry \cite{MS}. The maximal compact subgroup $C_U$ is denoted
 in the second column of Table 2. The  conjectured quantum symmetry is the
${\bf Z}$ valued subgroup of the classical symmetry (the
third column of Table 2).

\begin{table*}
\begin{tabular}{|l|c|c|c|}
D& Classical Duality-${ U}$ &
Maximal Compact Subgroup-${ C_U}$& Quantum Duality-${Q_U}$ \\
\hline
4 & $E_{{7(7)}}$&$SU(8)$ &$E_{{7(7)}}({\bf Z})$ \\
5 & $E_{{6(6)}}$&$USp(8)$ &$E_{{6(6)}}({\bf Z})$ \\
6 & $SO(5,5)$&$SO(5)\times SO(5)$ &$SO(5,5;{\bf Z})$ \\
7 & $SL(5,{\bf R})$&$SO(5)$ &$SL(5, {\bf Z})$ \\
8 & $SL(3,{\bf R})\times SL(2, {\bf R})$&$SO(3)\times U(1)$ &$SL(3, {\bf
Z})\times
SL(2,{\bf Z})$ \\
9 & $SL(2,{\bf R})\times {\bf R}^+$&$ U(1)$ &$SL(2, {\bf Z}) $\\
\end{tabular}
\caption[]{The classical   and  (conjectured)  quantum   duality
symmetries \cite{HT}  for toroidally
 compactified  Type  II string in $ 4\le D\le 9$. }
\end{table*}

\begin{table*}
\begin{tabular}{|l|c|c|c|}
D& Classical Duality-${ U}$ &
Maximal Compact Subgroup-${C_U}$& Quantum Duality-${ Q_U}$ \\
\hline
4 & $O(6,22)\times SL(2, {\bf R})$&$O(6)\times O(22)\times U(1)$ &$O(6,22;{\bf
Z})\times
SL(2,{\bf Z})$ \\
5 & $O(5,21)\times SO(1,1)$&$ O(5)\times O(21)$
 &$O(5,21;{\bf Z})\times {\bf Z}_2$ \\
6 & $O(4,20)\times SO(1,1)$&$O(4)\times
 O(20)$ &$O(4,20;{\bf Z})\times {\bf Z}_2$ \\
7 & $O(3,19)\times SO(1,1)$&$O(3)\times  O(19)$ &$O(3,19
;{\bf Z})\times {\bf Z}_2$ \\
8 & $O(2,18)\times SO(1,1)$&$ O(2)\times O(18)$
 &$O(2,18;{\bf Z})\times {\bf Z}_2$ \\
9 & $O(1,17)\times SO(1,1)$&$ O(17)$ &$O(1,17;{\bf Z})\times {\bf Z}_2$ \\
\end{tabular}
\caption[]{The classical and
 (conjectured) quantum  duality
symmetries of toroidally compactified  heterotic string \cite{MS} in $ 4\le D\le 9$.
}
\end{table*}

\subsubsection{Solution Generating Technique}

The general black hole solution is obtained by acting on a generating
solution   with the subset of $U$- ($T$-, $S$-) duality transformations.
First, the asymptotic value ${\cal
M}_{\infty}  $ of the scalar field metric
${\cal M}$ can be brought to the canonical value ${\cal M}_{0\, \infty}= {\bf
1}$ by a suitable $U$-duality transformation $\Omega_0$. The canonical value ${\cal
M}_{0\, \infty}$ is
preserved by
 $C_U$, {\it i.e.} the maximal compact subgroup of the $U$-duality group.
 The most general solution with the asymptotic behavior ${\cal
M}_{\infty} ={\cal
M}_{0\, \infty}$ is  then obtained by acting on the generating solution with a subset
of $C_U$ transformations, {\it i.e.} the $C_U$ orbits which are of
the form $C_U/C_0$ where $C_0$ is the subgroup preserving the generating
solution.  In particular, with this procedure 
the complete set of charges is obtained.  Indeed,  with the generating solution 
labeled by $n_0$ charges
($n_0=5,3,2$ for $D=4,5,\ge 6$,  respectively) and  the dimension of the $C_U$
orbits  being
$n_1$, then
$n_0+n_1$ is the correct dimension of the vector space of charges
 for the general solution.  Finally,  black holes with
arbitrary asymptotic values of scalar fields  ${\cal M}_{\infty}$ can then be
obtained
from these by acting
with $\Omega_0$.

\subsubsection{Charge Assignments for the Generating Solution}\label{cgs}
In the following we  shall list the charge  assignments  which were
 shown to be the charge assignments for the generating solutions
  for {\it both} toroidally compactified heterotic string \cite{CY}
  {\it and} Type II \cite{CH} string vacua.  
These charges are  chosen to be associated with the NS-NS sector of 
the Type II string (or toroidal sector of  heterotic string). 
Since these sectors are the {\it same} for both the heterotic and Type II string the
explicit form of the generating solution in both cases remains the same.  

\begin{itemize}
\item{ $D=4$} 

Generating  solutions are specified in terms of {\it five}
charge parameters. It is convenient to label these charges in terms of magnetic
$P^{(1,2)}_i$ and electric $Q^{(1,2)}_i$ charges associated with  $U(1)$ gauge fields 
$A_{\mu\,
i}^{(1,2)}$ of the Kaluza-Klein   (momentum) sector,
 and the  anti-symmetric tensor (winding) sector, respectively. Here, $i$ denotes the
$i$-th compactified direction.   The charge assignment of the generating solution is
the following:
$ Q_1^{(1)}$,  $Q_1^{(2)}$, $P_2^{(1)}$,
$P_2^{(2)}$  and
$q\equiv Q_2^{(1)}=-Q_2^{(2)}$.  
The generating
 solution   then carries five
charges   associated with the first two compactified toroidal
 directions of the NS-NS sector.

\item { $D=5$}

The generating solution is specified by three (electric)  charge
parameters:
$Q_1^{(1)}, \ Q_1^{(2)},$ and  $ {\tilde Q}$. Here $\tilde Q$ is the  electric charge 
of the gauge field, whose field
strength is related to the field strength of the two-form field $B_{\mu\nu}$  by a
duality transformation.

\item { $6\le D\le 9$}

The generating solution is  parameterized  by two
electric  charges:
$ Q_1^{(1)}, \ Q_1^{(2)}$. 

\end{itemize}

As an example we shall illustrate the  solution generating technique for
 the  case  of  $D=4$ toroidally compactified Type II string \cite{CH} 
 (the case  of $D=4$  toroidally compactified heterotic string  \cite{CY,CTII} 
 is analogous). 
One starts with the generating solution specified above by the five charge 
parameters.
The group $C_U=SU(8)$ preserves the canonical asymptotic values of the scalar fields
and only the  subgroup   $SO(4)_{L}\times SO(4)_{R}$
leaves the generating solution invariant.
Then acting with $SU(8)$ gives orbits
\begin{equation}
SU(8)/[SO(4)_L\times SO(4)_R]
\label{4du}
\end{equation}
of dimension  $63-6-6=51$. The $SU(8)$ action then  induces  $51$  new charge
parameters,
 which along with the original
five parameters  provide charge parameters for the general
 solution with 56 charges.

\subsection{Explicit Form of the Generating Solution}
In order to obtain the explicit form of the generating solution 
(which fully specifies  the $D$-dimensional space-time) one 
employs the following solution generating  technique. 
The  $D$-dimensional stationary solutions of  the theory are 
described by an  effective $(D-1)$-dimensional 
action, which  is obtained by  compactifying the $D$-dimensional 
 action of the theory along the Killing direction, {\it i.e.} time direction. Thus,
 acting with 
symmetry transformations of the $(D-1)$-dimensional 
 effective action on  a known stationary $D$-dimensional solution, one obtains 
 {\it new} $D$-dimensional stationary solutions. In particular,  by acting  with a
 subset of such  symmetry transformations on the 
$D$-dimensional   Kerr solution (neutral, rotating solution), 
the charged rotating  solutions are obtained.

Specifically, in the NS-NS sector of the theory the $(D-1)$-dimensional action 
 possesses $O(11-D,11-D)$  non-compact symmetry. 
  Acting with  boosts $SO(1,1)\subset 
O(11-D,11-D)$  \cite{SEN} on the $D$-dimensional 
Kerr solution, specified by the mass $m$ and $[{{D-1}\over 2}]$ 
angular momenta $l_{1,\cdots, [{{D-1}\over 2}]}$. 
{\it E.g.}, the two $SO(1,1)$ boosts with boost 
parameters $\delta_{1},\ \delta_{2}$ generate  the electric charges  
$Q^{(1),(2)}_1$ of the Kaluza-Klein $U(1)$ gauge field  
 and the two-form $U(1)$ gauge field, respectively.  The solution obtained in
that manner is specified by the ADM mass, {\it two} $U(1)$ charge 
parameters, and $[{{D-1}\over 2}]$ angular momenta 
$J_{1,\cdots, [{{D-1}\over 2}]}$. 
In  that manner the charged generating solution for black holes 
in $D\ge 6$ is obtained \cite{CYNear}. In $D=4,5$ the
 additional boost transformations in the NS-NS sector are needed (See Refs.
\cite{CY4s,CY5r}
 for more details.).

A program to obtain the explicit form of the generating solutions
 for general rotating black hole solutions  is close to completion.
Particular examples of solutions had been obtained in a number of papers (for
a recent review and references, see
 \cite{Horowitz}).
The explicit  expression for the  generating
solution  has  been obtained in
 $D=5$ \cite{CY5r}
and $D\ge 6$\cite{CYNear,Llatas}, however,
 in $D=4$ only  the five charge  static  generating solution
\cite{CY4s} (see also \cite{JMP})
 and the four
charge rotating solutions \cite{CY4r} were obtained.

\section{Bekenstein-Hawking Entropy}

While the explicit form of the  generating solution is complicated, 
the global space-time structure  can be easily deduced. In addition, 
 it turns out  that in this case the  Bekenstein-Hawking (BH)  entropy 
  $S_{BH}$ can be cast in a relatively simple, suggestive form.
Here, $S_{BH}=\textstyle{1\over 4G_N} A$, where $A$ is the surface area
determined at the outer-horizon 
$r_+$ and $G_N$ is Newton's constant (we follow  Ref. \cite{MP} for the 
definition of the ADM mass, charges and angular momenta and a convention that
the $D$-dimensional Newton's constant $G_N^D=(2\pi)^{D-4}/8$.). 
 
\subsection{$D=4$}
Here we quote results for the generating  rotating solution
  with four charges $ Q_1^{(1)}$,  $Q_1^{(2)}$, $P_2^{(1)}$,
$P_2^{(2)}$, only.  The global space-time is that of Kerr-Newmann 
black hole and the BH entropy can be cast in the following   suggestive  form
\cite{CY4r}:
\begin{eqnarray}
S_{BH}&=&16\pi [m^2(\prod^4_{i=1}\cosh\delta_i+\prod^4_{i=1}
\sinh\delta_i)+\cr
&&\{m^4(\prod^4_{i=1}\cosh\delta_i
-\prod^4_{i=1}\sinh\delta_i)^2-J^2\}^{1/2}], \cr
&&
\label{4dent}
\end{eqnarray}
where $m,\ l $ are  the ADM mass and the angular momentum 
per unit mass of the Kerr solution  
and 
$\delta_{1,2,3,4}$ are the four 
boosts specifying the four charges. 
The  ADM mass $M$, the four charges $Q^{(1),(2)}_1$, $P^{(1),(2)}_2$,
and the angular momentum $J$, of the rotating charged solution are defined in terms
of $m$,
$l$ and the four boosts as, respectively:
\begin{eqnarray}
M&=&4m({\rm cosh}^2 \delta_{1}+{\rm cosh}^2 \delta_{2}\cr
&&+{\rm cosh}^2 \delta_{3}+{\rm cosh}^2 \delta_{4})-8m, \cr
Q^{(1)}_1 &=&4m{\rm cosh}\delta_{1}{\rm sinh}\delta_{1},\ \ 
Q^{(2)}_1 = 4m{\rm cosh}\delta_{2}{\rm sinh}\delta_{2}, \cr
P^{(1)}_2 &=&4m{\rm cosh}\delta_{3}{\rm sinh}\delta_{3},\ \ 
P^{(2)}_2 = 4m{\rm cosh}\delta_{4}{\rm sinh}\delta_{4}, \cr
J&=&8lm({\rm cosh}\delta_{1}{\rm cosh}\delta_{2}{\rm cosh}\delta_{3}
{\rm cosh}\delta_{4}\cr
&&-{\rm sinh}\delta_{1}{\rm sinh}\delta_{2}
{\rm sinh}\delta_{3}{\rm sinh}\delta_{4}).  
\label{4dphys}
\end{eqnarray}
In the (regular) BPS-saturated limit ($m \to 0$, $l\to 0$, while
$m{\rm e}^{2\delta_{1,2,1,2}}$ are kept constant), the second  term in 
(\ref{4dent}) is zero and the BH entropy takes the form \cite{CY}:
\begin{equation}
S=32\pi m^2\prod^4_{i=1}\cosh\delta_i=
2\pi[Q^{(1)}_1Q^{(2)}_1P^{(1)}_2P^{(2)}_2]^{1\over 2}, 
\label{4dbpsent}
\end{equation}
 Note that as long as all the four charges of the 
 generating solution are non-zero the BH entropy (\ref{4dbpsent})
  is {\it non-zero} and the  black hole solution has the global
   space-time of the extreme Reissner-Nordstr\" om black hole.

Expression (\ref{4dbpsent}) 
has a straightforward 
generalization to the manifestly $S$- and $T$-duality invariant
form\cite{CTII}.
Namely, when expressed in terms of  28 conserved  quantized 
electric   charge  and  28 magnetic  charge lattice vectors 
$\vec{\alpha},\vec{\beta} \in \Lambda_{6,22}$ (of a toroidally 
compactified heterotic string), the surface area can be written  as \cite{CTII}
\begin{equation}
S=\pi[(\vec{\cal \alpha}^TL\vec{\cal \alpha})
(\vec{\cal \beta}^TL\vec{\cal \beta})-(\vec{\cal \alpha}^TL
\vec{\cal \beta})^2]^{1\over 2}, 
\label{4dvecexarea}
\end{equation}
where $L$ is the  $O(6,22)$ invariant matrix. In the  Type II case the result can be  expressed in terms
 of the unique $E_{7(7)}$ quartic  charge invariant \cite{KK,CH}.

Both, (\ref{4dvecexarea})  and the $U$-duality invariant form  of the BH entropy are
independent of asymptotic values 
of the moduli  and the dilaton coupling,  thus  suggesting
 that it may have a microscopic interpretation (as first anticipated by Larsen and
Wilczek
 \cite{LW}).

\subsection{D=5}
In this case the generating solution is specified by {\it three}
 charges. The global space-time is that of the five-dimensional Kerr-Newman 
solution. 
 The BH entropy can be cast \cite{CY4r} in the following suggestive form as a sum of the two
terms:
\begin{eqnarray}
S_{BH}&=&4\pi\Big\{[2m^3(\prod^3_{i=1}\cosh\delta_i+\prod^3_{i=1}
\sinh\delta_i)^2\cr
&&-\textstyle{1\over 16}(J_\phi-J_\psi)^2]^{1/2} \cr
& &+[2m^3(\prod^3_{i=1}\cosh\delta_i-\prod^3_{i=1} 
\sinh\delta_i)^2\cr
&&-\textstyle{1\over 16}(J_\phi+J_\psi)^2]^{1/2}
  \Big\}.
\label{5dent}
\end{eqnarray}
The ADM mass, the physical charges and the  two angular
 momentum components $J_{1,2}$  of the generating solution
  are related  to  the boost parameters $\delta_{1,2,3}$ as well as the
    ADM mass $m$ and two angular momentum components $l_{1,2}$ of the
five-dimensional
     Kerr solution in the following way:
\begin{eqnarray}
M&=&2m({\rm cosh}^2\delta_{1}+{\rm cosh}^2\delta_{2}+
{\rm cosh}^2\delta_{3})-3m,\cr
Q^{(1)}_1&=&2m{\rm cosh}\delta_{1}{\rm sinh}\delta_{1}, \cr
Q^{(2)}_1&=&2m{\rm cosh}\delta_{2}{\rm sinh}\delta_{2}, \cr
Q&=&2m{\rm cosh}\delta_{3}{\rm sinh}\delta_{3}, 
\cr
J_{\phi}&=&4m(l_1{\rm cosh}\delta_{1}{\rm cosh}\delta_{2}
{\rm cosh}\delta_{3}\cr
&&-l_2{\rm sinh}\delta_{1}{\rm sinh}\delta_{2}
{\rm sinh}\delta_{3}), \cr
J_{\psi}&=&4m(l_2{\rm cosh}\delta_{1}{\rm cosh}\delta_{2}
{\rm cosh}\delta_{3}\cr
&&-l_1{\rm sinh}\delta_{1}{\rm sinh}\delta_{2}
{\rm sinh}\delta_{3}).
\label{5ddef}
\end{eqnarray}
The  regular BPS-saturated limit (of the generating solution) is obtained by  
taking $m\to 0$, while keeping the three charges
$Q_1^{(1)}$, $Q^{(2)}_1$ and $Q$, as well as $J_\phi$ and 
$J_\psi$ finite. 
The surface area of the horizon is of the form:
\begin{equation}
S_{BH}=4\pi[Q^{(1)}_1Q^{(2)}_1Q-\textstyle{{1\over 4}}J^2]^{1\over 2}.
\label{5dbpsent}
\end{equation}
Expression (\ref{5dbpsent}) 
has a straightforward 
generalization to the manifestly $S$- and $T$-duality invariant
form\cite{CY5r}, expressed in terms of  26 
 conserved  quantized 
electric  charge lattice vectors
$\vec{\alpha} \in \Lambda_{5,21}$ (of toroidally 
compactified heterotic string) and one  conserved quantized charge $\tilde \beta$
(associated
with the gauge field related to the  two-form field by a duality transformation).
The surface area can be written  as \cite{CY5r}
\begin{equation}
S=4\pi[\textstyle{{1\over 2}}{\tilde \beta}(\vec{\cal \alpha}^TL\vec{\cal \alpha})
-\textstyle{{1\over 4}}J^2]^{1\over 2}. 
\label{5dareat}
\end{equation}
In the  Type II case the result can be  expressed in terms
 of the unique $E_{6(6)}$ cubic  charge invariant \cite{FK,CH}. Again,  in
 both the heterotic
 and Type II string case,  the five-dimensional BH entropy of 
 the BPS-saturated solution  does not depend
on the value of moduli or the gauge couplings.

\subsection{5$<$D$<$10}
The generating solution  has the global space-time structure 
of the  $D$-dimensional Kerr black hole. With all the angular
 momenta turned on the BH entropy can be expressed in a  
compact form only for four- and five-dimensional black holes 
\cite{CYNear} and  becomes progressively complicated as the 
dimensionality increases $D>5$.  

We shall therefore concentrate on the near BPS-saturated limit where the BH entropy
can
again be written in a compact form.
The near-BPS-saturated black holes have regular horizons  provided that the angular 
momentum parameters $l_{1,\cdots,[{{D-1}\over 2}]}$ have a 
magnitude which is smaller than that of $m$.  More precisely, in 
the near BPS-saturated limit
one has to take $m$ much smaller compared to ${\rm e}^{\delta_i}$, 
such that (when measured in units of $\alpha^{\prime}$) 
$Q_1^{(1),(2)}\gg m={\cal O}(1)$.  In addition, the angular 
momenta are kept small compared to charges, so that  $Q_1^{(1)}
Q_1^{(2)}\gg J^2_{1,\cdots,[{{D-1}\over 2}]}\gg 
{\sqrt{Q_1^{(1)}Q_1^{(2)}}}$.   The first inequality 
ensures the regular horizon, while the second inequality 
ensures that the contribution from angular momenta to 
the entropy is still non-negligible macroscopically.  Now, the
 BH entropy 
can be cast in the following form\cite{CYNear}:
\begin{eqnarray}
S_{BH}&=&2\pi \left[{4\over {(D-3)^2}}Q^{(1)}_1
Q^{(2)}_1(2m)^{2\over{D-3}}\right.\cr
&&\ \ \ \left.-{{2}\over{(D-3)}}
\sum_{i=1}^{[{{D-1}\over 2}]} J^2_i\right]^{1\over 2}. 
\label{nexarea}
\end{eqnarray} 

\section{Microscopic Entropy}
\subsection{Black Hole Micro-states as Elementary String Excitations}

The proposal to  identify the microscopic degrees of  electrically charged
 BPS-saturated black holes with those of  elementary string states  is 
due to  Sen 
\cite{SEN}. He proposed 
 to
 identify the 
  area of BPS-saturated electrically charged spherically 
symmetric solutions, evaluated at the {\it stretched} horizon,
with the  degeneracy of 
 BPS-saturated  states of string  theory (with the same charge assignment and
mass as
 that of the BPS-saturated electrically charged static black hole).  Since the 
stretched horizon is determined up to ${\cal O}
(\sqrt{\alpha^{\prime}})$, the identification of the two 
quantities agrees up to ${\cal O}(1)$, only.

For the {\it rotating}  electrically charged solutions one is faced with the problem
that
the BPS-saturated solutions have (in general) a naked singularity.
On the other hand, the area of rotating BPS-saturated  
black holes, evaluated at the stretched horizon (whose value  
is chosen to be independent of the angular momenta and electric 
charges), turns out to be {\it independent of the angular 
momenta}.  This result is therefore not in accordance with the 
expectations that BH entropy  should depend on the angular momenta, 
in order to be at least in qualitative agreement with the  
logarithm of the degeneracy of the corresponding
BPS-saturated string states with non-zero angular momenta. 

In  another proposal \cite{CYNear}, the area of the 
{\it near}-BPS-saturated  electrically charged black holes, 
 evaluated at their {\it true} horizon (\ref{nexarea}), 
should be identified  with the  degeneracy of states of the BPS-saturated  elementary
string
excitations with the {\it  same}  quantum numbers (for charges, angular momenta-spins
 and the mass).
Namely, the role of the stretched horizon of the BPS-saturated 
states is traded for the non-extremality parameter $m$ of the 
near-BPS-saturated states.  The degeneracy of these states
$d_{{\tilde N}_L}$  can be calculated \cite{CYNear} and yields:
\begin{eqnarray}
S_{stat} &\equiv& \log d_{{\tilde N}_L} \sim 
4\pi\sqrt{{\tilde N_L}}\cr
 &=& 4\pi\left(N_L - {1\over {2}} 
\sum_{i=1}^{[{{D-1}\over 2}]}J_i^2\right)^{1\over 2}. 
\label{statent}
\end{eqnarray}
In order to ensure the statistical nature of the entropy 
we need to maintain $N_L\gg {1\over {2}} 
\sum_{i=1}^{[{{D-1}\over 2}]}J_i^2$, while still allowing 
for the statistically significant contribution from spins, 
{\it i.e.} $\sqrt{N_L}\sum_{i=1}^{[{{D-1}\over 2}]}
J_i^2\gg 1$.

Note that in this case  the microscopic entropy (\ref{statent}) and the BH entropy 
(\ref{nexarea}) are in qualitative agreement (for $m={\cal O} (1)$).

\subsection{Black Hole Micro-states as Quantum Hair of Solitonic Strings}

An earlier  approach to calculate the 
microscopic entropy of four-dimensional BPS-saturated black holes 
  with regular horizons was initiated in Ref. \cite{LW} and was further elaborated on 
in Refs. \cite{CTII,T} and \cite{LWI}.  These approaches identify 
the microscopic black hole degrees of freedom as quantum hair 
\cite{LW} associated with particular small scale (marginal) 
perturbations \cite{CTII,LWI} of string theory, which do not 
change the large scale properties of the black hole solutions.  
On the one hand, the magnetic charges of such dyonic black 
hole solutions ensure that the classical solutions have regular 
horizons (and thus $\alpha^{\prime}$ corrections 
are under control), while on the other hand they effectively 
renormalize \cite{LW,CTII,T,LWI} the string tension of the 
underlying string theory.  Within this approach the correct 
 dependence  of statistical entropy on charges (and
angular
momenta) is obtained.
However,  in order to achieve a precise numerical agreement 
with  (\ref{4dbpsent})  and 
(\ref{5dbpsent})
additional assumptions on  relevant microscopic degrees of freedom are needed
\cite{T} (see also Ref. \cite{HMII} and references therein.).

\subsection{Black Hole Micro-states as D-branes}

Even though  we chose to parameterize the generating solutions 
 in  terms of fields of the toroidally compactified 
heterotic sting [or equivalently in terms of the NS-NS sector 
fields of the toroidally compactified Type IIA string], these 
 generating solutions  can be mapped, using 
string-string duality [or $U$-duality], onto configurations  
with R-R charges of a Type IIA string compactified on $K3\times T^2$ 
[or R-R charges of a Type IIA string compactified on $T^6$].  
Thus, they have a (higher-dimensional) interpretation in terms of the 
(intersecting) $D$-brane configurations and 
their microscopic degrees  are those of the  $D$-brane
configuration
(see Ref. \cite{Polchinski} for further details on the physics of $D$-branes.).

 The first agreement between  such a calculation of the microscopic entropy
and the BH one  was obtained by Strominger and Vafa \cite{SV}  in the case of static BPS-saturated  black
holes in $D=5$, and
 was further generalized to (near)-BPS-saturated (rotating) black holes in $D=5,4$ (for a
review see
\cite{Horowitz,Maldacena} and references therein.).  In spite  of the fact
that  the classical black hole picture is valid only when the D-brane system 
corresponds
to the strongly coupled  string theory (and thus the perturbative calculation 
need not be valid in this regime) the  numerical agreement between
 the the BH entropy and the perturbatively calculated result for the
statistical entropy is none the less  dramatic. It suggests that 
in the (near)-BPS limit the weak  string coupling calculations can 
be  reliably  carried over to the strong coupling regime.

We would also like to emphasize that the  BH entropy
 for  a generating solution of
 {\it non-extreme} rotating  black  holes  in $D=4$ (\ref{4dent})  and $D=5$  
 (\ref{5dent}) is cast in a suggestive form 
  as a {\it sum} of {\it two} terms of the type 
$\sqrt{n_L}$ and $\sqrt{n_R}$,  
the latter one disappearing in the BPS-saturated limit, or
extreme Kerr-Newman limit.  This structure strongly
suggests
 that  even  for   black hole solutions (far away from the extreme limit)
  there may be      an underlying microscopic description in 
  terms of excitations of  a {\it
 non-critical} string model (along the lines of a recent proposal of Horowitz and
 Polchinski\cite{HP}).

\section{Black Holes as Intersecting M-branes}
A  unifying treatment of string-theory 
 black hole properties  may
arise by identifying  such black holes as (toroidally) compactified
configurations of intersecting  two-branes  and five-branes of
eleven-dimensional M-theory. 
A discussion of  intersections  of certain
BPS-saturated  M-branes along with a proposal for intersection rules was
first given in \cite{PT}. A generalization to a number of
different harmonic functions specifying  intersecting
BPS-saturated M-branes  which  led to  a better understanding
 of these solutions and a   construction of 
new intersecting p-brane  solutions  in $D \leq 11$ was presented 
 in  \cite{TM} (see
also  related work \cite{KTT,DYA,DYB,CS,COS,GKT,BE,BL,PO}).
 Specific  configurations of that type reduce to the 
BPS-saturated black holes  with regular horizons 
in $D=5$ \cite{TM}  and $D=4$  \cite{KTT} 
 whose properties are determined by three and four charges (or 
harmonic functions), respectively.

One can further generalize such BPS-saturated  intersecting $M$-branes to the case of 
{\it non-extreme  static} intersecting $M$-branes \cite{CT} 
as well as  {\it non-extreme  rotating} 
intersecting $M$-brane solutions \cite{CYRM}.  
 Such configurations should be interpreted as {\it bound state} 
solutions of $M$-branes with a {\it common} non-extremality parameter 
and {\it common} rotational parameters associated with the transverse 
spatial directions of the $M$-brane configuration.

There exists   a general algorithm for constructing 
the overall conformal factor and internal components of the
eleven-dimensional metric for such configurations. 
The space-time describing the {\it internal} part of such (intersecting) 
configurations is specified {\it entirely} by ``harmonic functions'' for each 
constituent $M$-brane (associated with each charge source) and the   
``non-extremality functions'' (associated with the Kerr mass), 
which are  in general  modified by functions that 
depend on  the rotational parameters.  

On the other hand, the transverse part of the configuration, which 
reflects the axial symmetry of the solution, involves charge sources 
as well as the  rotational parameters in a more involved manner and cannot 
be simply written in terms of modified harmonic functions and non-extremality 
functions, only. However, in the case of 
static solutions the transverse 
part has a uniform structure of the form $f^{-1}(r)dr^2+r^2d\Omega^2_{D-1}$ 
with $d\Omega^2_{D-1}$ given by the infinitesimal length element of the 
unit $(D-2)$-sphere $S^{D-2}$.

As an explicit example,  we write down 
a  structure for the 
intersection of two membranes and two five-branes $(2\perp 2\perp 5\perp 5)$, 
which becomes, after a dimensional reduction, a
four-dimensional non-extreme black hole with four-charges.  
Such an intersecting $M$-brane solution has the following 
structure for the eleven-dimensional metric\cite{CT,CYRM}:  
\begin{eqnarray}
ds^2_{11}&=&(T_1T_2)^{-1/3}(F_1F_2)^{-2/3}\left[
-T_1T_2F_1F_2fdt^2 \right.
\cr
&+&F_1(T_1dy^2_1+T_2dy^2_3)+F_2(T_1dy^2_2+T_2dy^2_4)\cr
&+&F_1F_2(dy^2_5+dy^2_6+dy^2_7)\cr
&+&\left.f^{\prime\,-1}dr^2+r^2d\Omega^2_3\right]
\label{4intmbr}
\end{eqnarray}
where the ``modified'' harmonic functions $T_i$ and $F_i$ are associated, 
respectively, with the electric charges $Q_i=2m\cosh\delta_{ei}
\sinh\delta_{ei}$ and the magnetic charges $P_i=2m\cosh\delta_{pi}
\sinh\delta_{pi}$, and the non-extremality functions$f,f'$  are given by
\cite{CYRM} 
\begin{eqnarray}
T^{-1}_i&\equiv&1+f_D{{2m\sinh^2\delta_{ei}}\over r}, \cr
F^{-1}_i&\equiv& 1+f_D{{2m\sinh^2\delta_{pi}}\over r}, \ \ \ \ \ \ i=1,2,  
\cr
f&\equiv&1-f_D{{2m}\over r},\cr
f^{\prime}&\equiv& f_D\left(1+{{l^2}\over{r^2}}-{{2m}\over r}\right), 
\label{4intharm}
\end{eqnarray}
where $f^{-1}_D=1+{{l^2\cos^2\theta}\over{r^2}}$ and $l$ is the angular
momentum of the neutral solution.  
Here, the  expression for $d\Omega^2_3$, 
  reduces in the case of static
solution
($l=0$) to the  line element of $S^2$  (in the case of non-zero $l$ 
the explicit expression  for $d\Omega_3^2$ is not known). For the  explicit solutions 
  of    one, two, and three intersecting   $M$-branes  see
Ref. \cite{CT}  (in the case of static solutions)  and  Ref. \cite{CYRM} (in the case of 
 rotating solutions).

\section{Conclusions}

 Over the last year  dramatic  progress in
 understanding  classical  and microscopic properties of black holes in string
theory has been made. 
 In this contribution we have addressed some aspects of this progress.
 
 In particular, we have
 reviewed the  structure of the  Bekenstein-Hawking entropy of  general rotating
black hole
 solutions in toroidally compactified string theory (in diverse dimensions)
  and  have emphasized  connections of its  suggestive structure 
   to its microscopic
  origin (in the (near)-BPS
limit) from the point of view of elementary string excitations, 
 small-scale oscillations of an  underlying higher-dimensional configuration, as well 
as  D-brane  configurations.
Even more tantalizing is the fact that   generating solutions for  
four- and five-dimensional
{\it  non-extreme }  black holes  still possess 
 a suggestive  form, indicating that even in this case 
    there may be 
     an underlying microscopic description in terms of excitations of   {\it
 non-critical} string  configurations.

  The suggestive interpretation of the  these configurations as {\it  non-extreme 
  rotating} $M$-brane configurations may  also  turn out to provide a useful tool to
  address  properties of the underlying black holes.
 
\section{Acknowledgments}
I  would like to thank  C. Hull,  A. Tseytlin,  and especially D. Youm
 for fruitful collaborations and discussions on topics presented in this
contribution.
  The work is supported by the U.S. DOE Grant No. DOE-EY-76-02-3071, the NATO
collaborative research grant CGR No. 940870 and the National
Science Foundation Career Advancement Award No. PHY95-12732.

\end{document}